# SPICE Simulation of tunnel FET aiming at 32 kHz crystal-oscillator operation

**Tetsufumi Tanamoto[1], Chika Tanaka[1], Satoshi Takaya[1], and Masato Koyama[1]**

[1]*Corporate R and D Center, Toshiba Corporation, 1, Komukai Toshiba-cho, Saiwai-ku, Kawasaki 212-8582, Japan*

**Abstract:** We numerically investigate the possibility of using Tunnel field-effect transistor (TFET) in a 32 kHz crystal oscillator circuit to reduce power consumption. A simulation using SPICE (Simulation Program with Integrated Circuit Emphasis) is carried out based on a conventional CMOS transistor model. It is shown that the power consumption of TFET is one-tenth that of conventional low-power CMOS.
**Classification:** Electron devices, circuits and modules (silicon, compound semiconductor, organic and novel materials)

## 1 Introduction

In the coming era of the Internet of Thing (IoT), low-power operation is a key factor in device development. Moreover, it is considered that the end of transistor scaling is inevitable in the near future. Given this background, the development of the tunnel field-effect transistor (TFET) has been attracting great interest [1-8]. TFET's steep subthreshold-slope (SS) current-voltage characteristics would enable low-power operation in many applications.

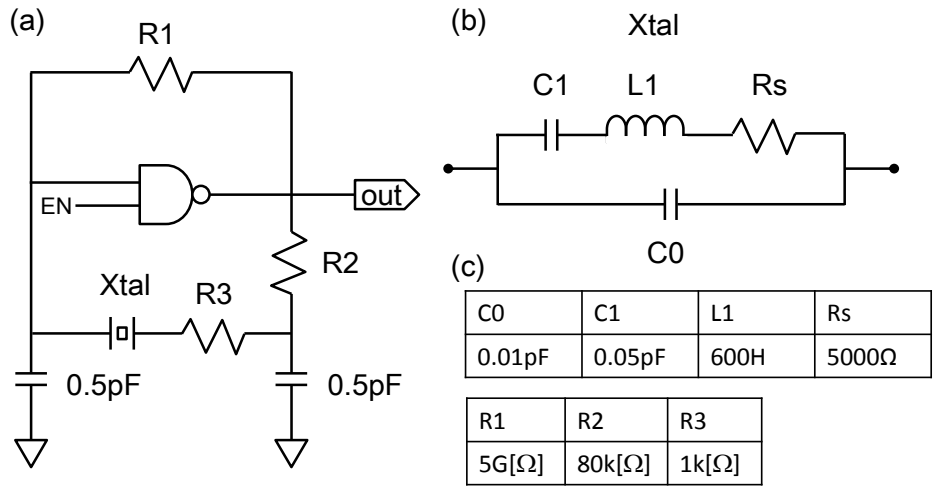

**Fig. 1.** (a) Crystal-oscillator circuit. (b) Equivalent circuit of quartz crystal using four elements. (c) Simulation parameters used in the oscillating circuits.

A crystal oscillator is implemented in most digital circuits to generate clock frequency. Crystal oscillators convert a direct current from a power supply to a periodically oscillating current signal. A frequency of 32 kHz is commonly used in many applications to generate a real time clock. This is because 32 kHz is a power of 2 ($32768=2^{15}$) value, and a precise 1 second period (1 Hz frequency) is obtained by using a 15 stage binary counter. Because real-time clocks are always working in IoT devices such as a smart phone or a smart watch, the power reduction of the 32 kHz crystal oscillator circuits leads directly to longer device usage. Here, we show SPICE (Simulation Program with Integrated Circuit Emphasis) simulation results of oscillator circuit using TFET based on a compact model [9,10]. We applied TFET in a 32 kHz crystal oscillator circuit.

In real applications, crystal oscillator circuits have various forms including such as amplification circuits depending on their target devices. Here, we simulate the basic crystal oscillator circuits shown in Fig.1. Although this circuit is very simple, fundamental performance can be investigated by this circuit. The crystal part that determines the frequency of the circuit is represented by its equivalent circuit (Fig.1 (b)). There are four transistors in the NAND gate that works as an amplifier of the resonant signal. We numerically compare the circuit performance in which the NAND gate consists of four TFETs with that of conventional CMOSs. We use HSPICE simulator and transistor model based on 65 nm CMOS parameter set (comparison is carried out by transistors of the gate length of 120nm). Even a small difference of transistors in the NAND gate can induce a large difference in circuit performance, because the crystal-oscillation repeats the charging and discharging of transistors. We show that the replacement of four conventional transistors by TFETs in the NAND gate results in a large reduction in power consumption.

|  | TFET1 | TFET2 | CMOS1 | CMOS2 |
|---|---|---|---|---|
| $I_{on}/I_{off}$ | 1.0E+7 | 8.0E+6 | 1.0E+8 | 5.0E+6 |
| $V_{th}$[V] | 0.287 | 0.342 | 0.285 | 0.331 |

**Table. 1.** Four type of transistors that are simulated in this paper. CMOS1 and CMOS2 are two types of low-powered CMOSs. The model parameters of TFET2 are abstracted from Ref. [8]. TFET1 is improved version of TFET2.

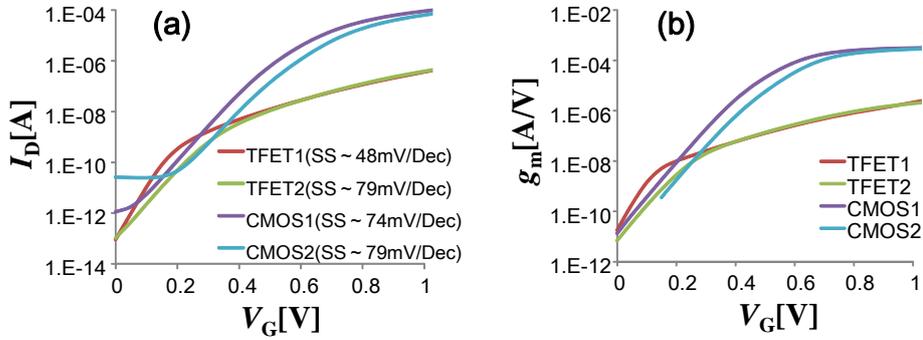

**Fig. 2.** (a) Simulated $I_d$-$V_g$ characteristics of four transistors of Table 1. (b) Simulated $g_m$ characteristics of four transistors. The subthreshold-slope (SS) differs depending on each device.

.

## 2 Basic characteristics of simulated TFET devices

We compare the two types of TFET models ($L$=120nm) with two types low-power CMOS models ($L$=120nm) shown in Table 1. TFET1 is the ideal version model of experimentally fabricated TFET2 of Ref.[8]. Because TFET2 has not yet achieved the low SS (〜79mV/Dec), the SPICE model parameters of TFET1 are adjusted to achieve lower SS(〜48mV/Dec). The CMOS1 and CMOS2 are based on the 65 nm CMOS transistor models aiming at low power operations, and CMOS2 has relatively higher $V_{th}$ than CMOS1 aiming at less low-power operation. The $I_d$-$V_g$ and $g_m$ characteristics at room temperature are shown in Fig.2 (a-b). The $g_m$ is a transconductance of the transistor and contributes to the gain of the amplification circuits.

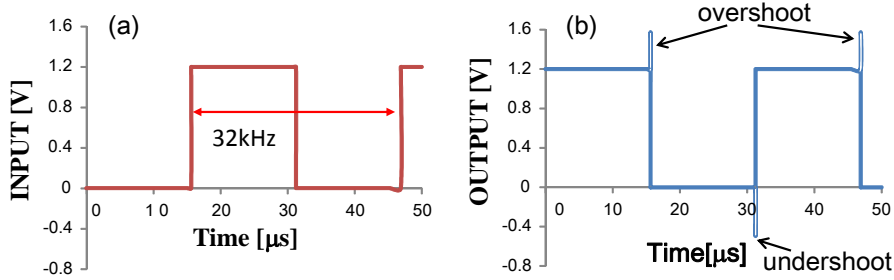

**Fig. 3.** The output of an inverter to 32kHz input pulse. Overshoot and undershoot as a result of Miller effect. (a) Input. (b) Output.

**3**

## Simulation results

3.1 Voltage overshoot and undershoot

  The disadvantage of TFET is that the Miller effect of TFET is about twice that of conventional MOSFET [11], which is undesirable in conventional digital circuits. The source side tunneling barrier in the TFET structure enhances the Miller capacitance, resulting in large voltage overshoot/undershoot in its switching performance. Figure 3 shows a simulated waveform of an inverter to a 32 kHz input pulse. We can see a large overshoot and undershoot. It can be also seen that the widths of overshoot and undershoot terminate within 40 ps. Although it is considered the overshoot/undershoot restrict the performance of TFETs, in the next section, we will show that the overshoot and undershoot help to enhance the amplitude of periodic oscillations in the present application.

3.2 Results of crystal oscillator circuits

  Although the present circuit (Fig. 1) has only four transistors in its NAND gate, the oscillation leads to repetitions of charging and discharging, resulting in clarifying the difference between TFET and conventional CMOS. The resistances R1, R2 and R3 determine the oscillation performance. R1 is a large resistance to prevent the short of input and output of NAND gate. R2 is a feedback resistor of the amplification. R3 contributes to the oscillation margin, and should be five times larger than that of Rs in Fig. 2 (b). Because Rs=5 k$\Omega$, the oscillation for R3>25 k$\Omega$ is desirable.

    Figures 4 (a-b) show the gain and the phase of the circuits using TFET1, respectively. Figure 5 shows a typical 32 kHz oscillation waveform. It can be seen

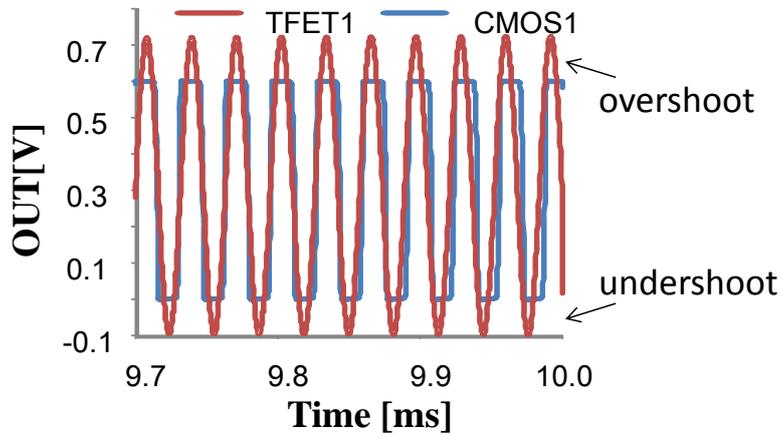

**Fig. 5.** Part of the output of 32 kHz crystal-oscillator of TFET1 for $V_{DD}$=0.6[V]. We can see the effect of overshoot and undershoot, which help to enhance the amplitude of the periodic oscillations.

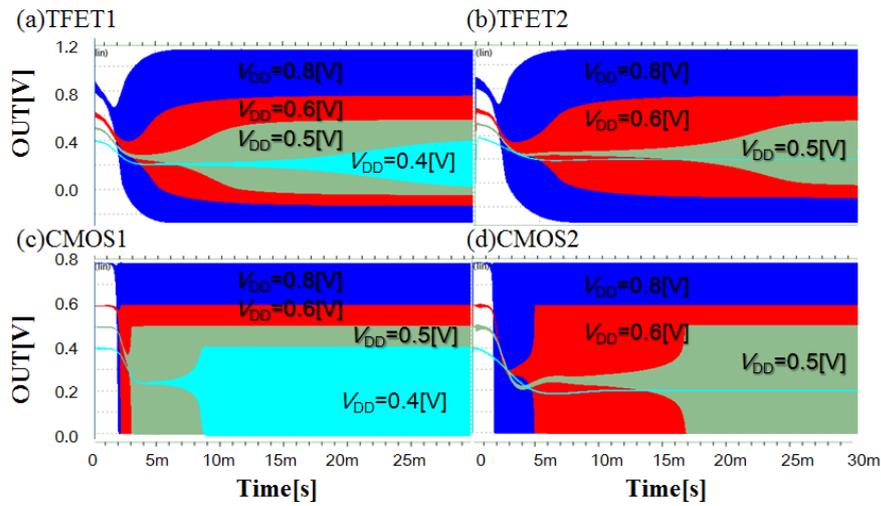

**Fig. 6.** Output of 32 kHz Oscillation when $V_{DD}$ is changed. (a) TFET1, (b) TFET2, (c) CMOS1, (d) CMOS2. TFET1 and CMOS1 show clear oscillation at $V_{DD}$=0.4 [V] while TFET2 and CMOS2 show no clear oscillation at $V_{DD}$=0.4[V]

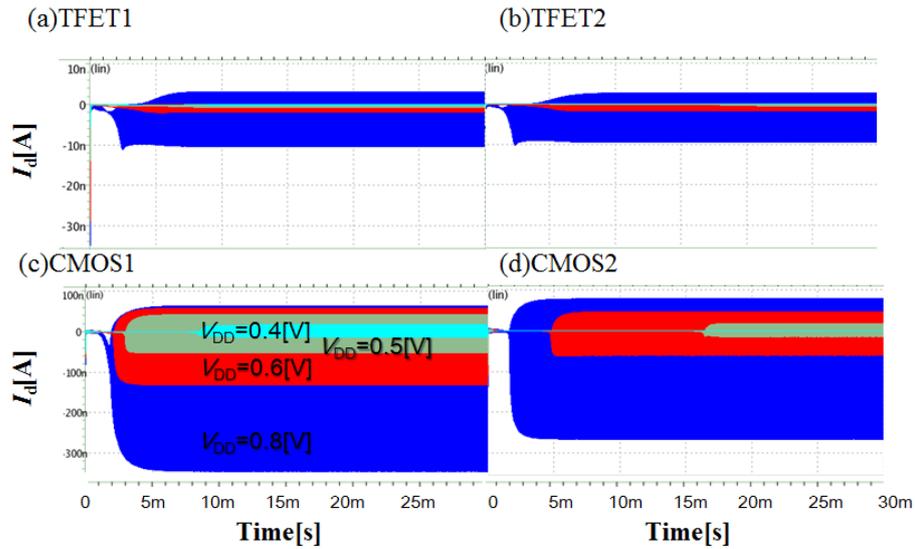

**Fig. 7.** Current of 32 kHz Oscillation when $V_{DD}$ is changed. (a) TFET1, (b) TFET2, (c) CMOS1, (d) CMOS2. The power consumption is calculated from these currents. Thus, it can be seen that the power consumptions of TFETs are much smaller than those of CMOSs.

## 4 Conclusions

SPICE simulation of a 32 kHz crystal oscillator circuit is carried out using a TFET compact model. It is shown that the power consumption of the circuit using TFET is much smaller than that of conventional CMOS. It is also found that the large Miller effect due to the TFET structure works positively for generating the oscillating waveforms. Because the 32 kHz crystal oscillator circuit is widely used in IoT devices, the application of TFET is proven to be effective.

| $V_{DD}$ | TFET1 | TFET2 | CMOS1 | CMOS2 |
|---|---|---|---|---|
| 0.8[V] | 8.883e-10[W] | 7.270e-10[W] | 9.981e-09[W] | 5.659e-09[W] |
| 0.4[V] | 4.274e-11[W] | NG | 1.959e-10[W] | NG |

**Acknowledgments**

This work was supported by JST CREST Grant Number JPMJCR1332, Japan. We thank S. Kawanaka, H. Hara, K. Kushida, K. Adachi, M. Goto, M. Fujimatsu, A. Nishiyama and S. Yasuda for discussions.